\documentclass[12pt]{article}
\usepackage{epsfig}
\usepackage{url}
\makeatletter
\def\alt{\mathrel{\mathpalette\gl@align<}}
\def\agt{\mathrel{\mathpalette\gl@align>}}
\def\gl@align#1#2{\lower.6ex\vbox{\baselineskip\z@skip\lineskip\z@
\ialign{$\m@th#1\hfil##\hfil$\crcr#2\crcr\sim\crcr}}}
\makeatother

\begin{document}
\begin{flushright}
{\tt hep-ph/0611268}\\
MIFP-06-31 \\
November, 2006
\end{flushright}
\vspace*{2cm}
\begin{center}
\baselineskip 20pt
{\Large\bf
Modification of the Unitarity Relation for $\sin2\beta$-$V_{ub}$\\
in Supersymmetric Models } \vspace{1cm}

{\large
Bhaskar Dutta and Yukihiro Mimura}
\vspace{.5cm}

{\it
Department of Physics, Texas A\&M University,
College Station, TX 77843-4242, USA
}
\vspace{.5cm}

\vspace{1.5cm}
{\bf Abstract}
\end{center}
Recently, a more than
2$\sigma$ discrepancy has been observed between the well measured
inclusive value of $V_{ub}$ and the predicted value of $V_{ub}$ from
the  unitarity triangle fit using the world average value of
$\sin2\beta$. We attempt to resolve this tension in the context of
 grand unified SO(10) and SU(5) models where the neutrino mixing matrix is responsible for flavor
 changing neutral current at the weak scale and the models with
non-proportional $A$-terms (can be realized simply in the context of intersecting D-brane models)
and investigate the interplay between the constraints arising from
 $B_{s,d}$-$\bar B_{s,d}$ mixings, $\epsilon_K$, Br($\tau\to\mu\gamma$), Br($\mu\to e \gamma$)
and a fit of this new discrepancy. We also show that the ongoing
measurement of the phase of  $B_s$ mixing will be able to identify
the grand unified model. The measurement of Br($\tau\to e\gamma$)
 will also be able to test these scenarios,
especially the models with non-proportional $A$-terms.

\thispagestyle{empty}

\bigskip
\newpage

\addtocounter{page}{-1}

\section{Introduction}
\baselineskip 18pt

Recent measurement of $B_s$-$\bar B_s$ oscillation
\cite{Abulencia:2006ze}
not only can examine the Kobayashi-Maskawa theory
\cite{Kobayashi:1973fv}, but also
 can probe the existence of new physics such as supersymmetric (SUSY) models.
Accurate measurements of the mass differences for $B_d$-$\bar B_d$
and $B_s$-$\bar B_s$, time dependent CP asymmetry of $B_d \to J/\psi
\,K$ decay \cite{Bigi:1981qs}, and Cabibbo mixing can determine the
unitarity triangle.
Since one angle and two sides are determined from these
measurements, one can predict the remaining  angles and the  side.
The experimental measurements of the remaining two angles still have
large errors, while the measurement of the remaining side (naively
$|V_{ub}/V_{cb}|$) has become accurate recently, especially from
inclusive $B$ decay data. 
One can thus compare the experimental measurement of $|V_{ub}|$ 
with the value from the global fit of the 
CKM parameters \cite{Yao:2006px,Bona:2006sa,Charles:2004jd}
and 
it turns out that there exists a more
than  2$\sigma$ discrepancy between the experimental measurement and
the unitarity prediction of $|V_{ub}/V_{cb}|$
\cite{Yao:2006px,Bona:2006sa}. 
This discrepancy may
be an indication of new physics.

The SUSY models are the most attractive candidates for new physics.
The gauge hierarchy problem can be solved and a natural aspect of
the theory  can be developed from the weak scale to the ultra high
energy scale.
In fact, the gauge coupling constants in the standard model (SM) can
unify at a high scale using the  renormalization group equations
(RGEs) involving the particle contents of the minimal SUSY standard
model (MSSM), which indicates the existence of grand unified
theories (GUTs).
While the experimental result for gauge unification is successfully
explained in the gauge sector, the flavor sector remains a problem
in SUSY theories. The SUSY extension of the SM needs SUSY breaking
terms, and there can be a lot of parameters in the flavor sector in
general. The SUSY breaking scalar masses in general can induce large
flavor changing neutral currents (FCNCs), which contradict
experimental results. In order to avoid the SUSY FCNCs, the masses
of  SUSY scalar particles (squarks and sleptons) need to be
degenerate especially for the first two generations as long as the
SUSY particles are lighter than 2-3 TeV \cite{Gabbiani:1988rb}.

The flavor degeneracy of SUSY breaking scalar masses is often
assumed at a cutoff scale such as the GUT scale or the Planck scale.
The degeneracy can be realized when the K\"ahler metric is universal
in   flavor indices. However, even if the degeneracy is realized at
a scale, the RGE evolution still can induce flavor violation for
squarks and sleptons at low energy. The flavor violation induced by
RGEs is suppressed by loop factors, and thus, can satisfy the
current FCNC experimental bounds. The ongoing and future experiments
may reveal the existence of small flavor violation, and therefore
the footprint of GUT or Planck physics may be obtained since the
pattern of flavor violation depends on the unification of matters,
contents of heavy particles, and the nature of SUSY breaking.

In the scenarios where a small amount of flavor violation is
generated at low energy through RGEs, the seeds of flavor violation
are implanted in the Yukawa couplings and SUSY breaking scalar
trilinear couplings ($A$-terms).
The up- and down-type Yukawa matrices are not diagonalized
simultaneously, and thus, flavor mixings are generated in charged
current interaction. The FCNCs via RGEs can originate from the
mixing matrices characterized by the CKM (Cabibbo-Kobayashi-Maskawa)
and the MNSP (Maki-Nakagawa-Sakata-Pontecorvo) matrices for quarks
and leptons, respectively.
In the MSSM, the induced FCNCs in the quark sector are not large
since the CKM mixings are small. On the other hand,  sizable FCNC
effects can be generated in the lepton sector since two neutrino
mixings in MNSP matrix are large due to the
 observation of solar and atmospheric neutrino oscillations.
Thus, a testable amount of flavor violating lepton decay can be
observed \cite{Borzumati:1986qx}. In grand unified models, as a
consequence of the quark-lepton unification, the large neutrino
mixings can also generate flavor violation in the quark sector
\cite{Barbieri:1994pv}.

There could be another source of flavor violation. In the minimal
supergravity mediated SUSY breaking, the $A$-term couplings are
proportional to the Yukawa couplings, and therefore the effects are
already discussed above. In general, the $A$-term couplings can
possess non-proportional terms which can be new sources of flavor
violation. This can happen even if the K\"ahler metric is flavor
universal.
Actually, the $A$-term includes the following three parts: 1)
universal term which originates from local SUSY breaking, 2)
non-universal term which comes from non-trivial K\"ahler connection
(namely, non-canonical K\"ahler metric), 3) non-proportional term
which is generated when the Yukawa coupling is a function of moduli
fields which acquire non-zero $F$-components.
The $A$-term coupling in  first and second parts
 can be proportional to the Yukawa coupling when
the K\"ahler metric is flavor universal to make the SUSY breaking
scalar mass degenerate. In the third part, the  $A$-term couplings
are proportional to the derivative of the Yukawa couplings in moduli
fields, and the derivative of the Yukawa matrix is not necessarily
proportional to the Yukawa coupling itself.
%
%
%
Thus, it is possible that $A$-terms contain the seeds of flavor
violation  which does not depend on the CKM and MNSP mixings, even
if the SUSY breaking masses are degenerate at the string scale.

In this paper, we investigate the prediction of flavor violation
when the  unitarity relation of $V_{ub}$ and CP asymmetry of $B_d
\to J/\psi K$ is modified by the SUSY FCNC contribution.
We assume that the SUSY breaking scalar masses are degenerate at the
unification scale due to the nature of K\"ahler metric of the matter
sector. The sources of flavor violation induced via RGEs can be
the Yukawa couplings and the $A$-terms.
We will consider the following two typical cases when  the FCNCs are
induced in SUSY breaking scalar mass matrices via RGEs. In the first
case, the FCNCs originate from large neutrino mixings, and in the
second case, the scalar trilinear couplings are not proportional to
Yukawa couplings.

In the first case, we consider the SU(5) and SO(10) grand unified
models. In a SU(5) model, the right-handed down-type quarks and the
left-handed lepton doublet are unified in one multiplet. Thus, a
sizable flavor violation can be generated in the SUSY breaking mass
matrix for the right-handed down-type squarks. In a SO(10) model,
all the  matter fields are unified in one multiplet, and thus, both
left- and right-handed squark mass matrices can have sizable flavor
mixing. We will see that  SO(10) models can have larger FCNC effects
in $B$-$\bar B$ mixings compared to  SU(5) models.

In the second case, we will investigate the intersecting D-brane
model \cite{Berkooz:1996km,Blumenhagen:2005mu} where the
non-proportional $A$-terms have been realized explicitly. In this
scenario, the  SUSY breaking scalar masses are degenerate
\cite{Camara:2003ku}. In a simple intersecting D-brane model, the
Yukawa matrices are rank 1 matrices plus small corrections
\cite{Cremades:2003qj,Dutta:2005bb}, and consequently, the
non-proportional part of the $A$-term has a simple structure
\cite{Dutta:2006bp}.

In both cases, the important constraints come from flavor violating
lepton decays, $\mu \to e\gamma$ and $\tau\to\mu\gamma$, mass
difference of $B_s$-$\bar B_s$, and CP violation in $K$-$\bar K$
mixing. We will see that the experimental constraints restrict the
parameters in neutrino mixings and the SUSY particle spectrum. Once
the discrepancy between the unitarity prediction and experimental
measurements is explained by the SUSY contribution, the phase of
$B_s$-$\bar B_s$ mixings \cite{Blanke:2006ig,Dutta:2006gq} and
flavor violating lepton decays in these models get constrained. We
will investigate the predictions of the models. It is interesting to
note that the phase of  $B_s$-$\bar B_s$
 mixing is being measured by
the time dependent CP asymmetry of $B_s \to J/\psi \phi$ decay and
the CP asymmetry of semileptonic $B_s$ decay, and we already have a
value for this phase  from D\O\ \cite{D0}.

This paper is organized as follows. In section 2, we will go through
the unitarity relation of $V_{ub}$ and CP asymmetry of $B_d \to
J/\psi K$ to understand the  2$\sigma$ discrepancy between
 the unitarity prediction
and experimental measurement of $|V_{ub}|$.
In section 3,
 we will study the origin of flavor violation from the neutrino mixings in GUT models,
 and investigate the constraint and the consequences
 of the modification of unitarity relation by SUSY contributions.
In section 4,
 we will investigate the non-proportional $A$-term
 in the context of intersecting D-brane models,
 and see the prediction of the models when it explains the 2$\sigma$.
Section 5 is devoted to conclusions and discussions.

\section{Unitarity prediction of $|V_{ub}|$}

The recent accurate measurement of $B_s$-$\bar B_s$ mass difference,
as well as $B_d$-$\bar B_d$, can determine one side of the unitarity
triangle. In addition, one angle  obtained from the   CP asymmetry
of $B_d \to J/\psi K$ and one side of the unitarity triangle
determined by the Cabibbo mixing are accurate quantities. Therefore, the unitarity
triangle can be determined accurately. The remaining  side and the
angles are predicted if there is no effect from new physics.

The unitarity of the CKM matrix gives rise to the following
equation. The unitarity condition,
\begin{equation}
V_{ud} V_{ub}^* + V_{cd} V_{cb}^* + V_{td} V_{tb}^* =0,
\label{unitary}
\end{equation}
can be rewritten to obtain the following relation,
\begin{equation}
|V_{ud} V_{ub}^*|^2 = |V_{cd}V_{cb}^*|^2 + |V_{td}V_{tb}^*|^2
- 2 |V_{cd}V_{cb}^*V_{td}V_{tb}^*| \cos \beta \,,
\end{equation}
where
$\beta = \phi_1 \equiv {\rm arg}\, ({V_{td} V_{tb}^*}/{V_{cd} V_{cb}^*})\,.$
%
%
Using approximate relations, $|V_{cb}|\simeq |V_{ts}|$ and $V_{tb} \simeq 1$,
we obtain
\begin{equation}
\left|V_{ub}\right|
\simeq
\left|\frac{V_{cb}}{V_{ud}}\right|
\sqrt{|V_{cd}|^2+ \left|\frac{V_{td}}{V_{ts}}\right|^2
-2 \left|V_{cd}\frac{V_{td}}{V_{ts}}  \right| \cos \beta}\,.
\end{equation}
The recent measurement of $B_s$-$\bar B_s$ oscillation indicates
$|V_{td}/V_{ts}| = 0.206^{+0.008}_{-0.006}$ \cite{Abulencia:2006ze}.
Using world average of $\sin2\beta$ obtained from
 $B_d \to J/\psi K$ \cite{unknown:2006aq,Barberio:2006bi},
\begin{equation}
\sin2\beta = 0.674 \pm 0.026,
\end{equation}
$|V_{cd}| = 0.2258$, 
and
$|V_{cb}| = (41.6 \pm 0.6) \times 10^{-3}$ \cite{Yao:2006px},
we obtain
\begin{equation}
|V_{ub}| = (3.49\pm 0.17) \times 10^{-3} \ (\mbox{unitarity}).
\end{equation}
The $|V_{ub}|$ obtained above  can be compared to the combined
data analysis
$|V_{ub}|=(3.5\pm 0.18) \times 10^{-3}$ obtained by UTfit and
CKMfitter~\cite{Bona:2006sa,Charles:2004jd,Ahuja:2006fv,ball}.
We note that the $|V_{ub}|$ prediction from the unitarity triangle is
insensitive to $|V_{td}/V_{ts}|$ error since $\alpha = \phi_2 \equiv
{\rm arg}\, (V_{ud}V_{td}^*/V_{td}V_{tb}^*)$ is close to 90$^{\rm o}$. The
prediction for $\alpha$ from the unitarity triangle using two sides
$|V_{cd}|$, $|V_{td}/V_{ts}|$ and an angle $\beta$ is $\alpha =
(93.2\pm 5.7)^{\rm o}$.
The $\sin2\beta$-$V_{ub}$ relation is plotted in Figure 1.

%
The experimental measurement of $|V_{ub}|$ \cite{Yao:2006px,Barberio:2006bi}
is
\begin{eqnarray}
|V_{ub}| &=& (4.49 \pm 0.19 \pm 0.27)\times 10^{-3} \ (\mbox{inclusive}), \\
|V_{ub}| &=& (3.84^{+0.67}_{-0.49})\times 10^{-3} \ (\mbox{exclusive}).
\end{eqnarray}
The tension between the experimental data and the unitarity prediction
of $|V_{ub}|$ is significant especially for  the inclusive $B$ decay data
 which has less theoretical error compared to the exclusive case.
The unitarity condition does not agree with the inclusive data
within 99\% confidence level.
The data from the exclusive decay $B \to \pi \ell \bar \nu$ includes
large uncertainty from lattice calculation. The statistical error of
the inclusive data is reduced in recent experiments at Babar and
Belle~\cite{Barberio:2006bi,Kowalewski}, and the inclusive data has
become more reliable.

\begin{figure}[tb]
 \center
 \includegraphics[viewport = 20 20 280 225,width=8.5cm]{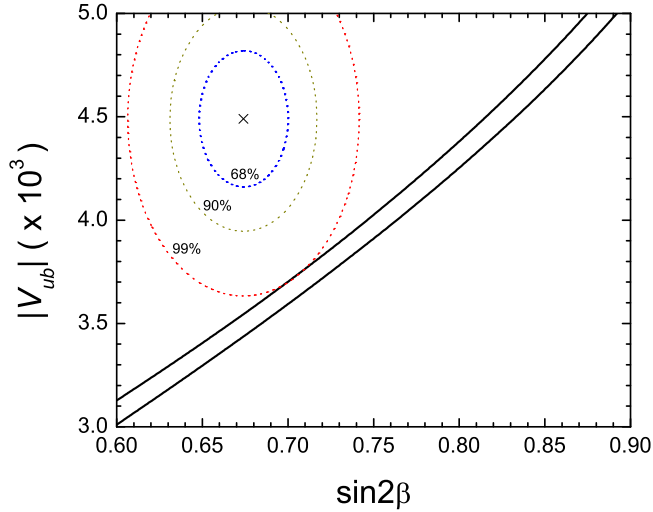}
 \caption{The 1$\sigma$ range of the unitarity relation between
 $|V_{ub}|$ and $\sin2\beta$ (solid lines).
 The central value of experimental measurements is plotted at x.
 We also plot the experimental region for 68\%, 90\%, and 99\% confidence levels
 assuming that the errors are Gaussian.
 }
\end{figure}

The discrepancy between the unitarity condition and the experimental
measurement may indicate new physics. We consider a scenario in
which the unitarity condition, Eq.(\ref{unitary}), is really true,
but the SUSY contribution modifies the $\sin2\beta$ measurement by
$B_d \to J/\psi K$. We will study different  types of boundary conditions
(SU(5), SO(10) GUTs or non-proportional $A$-terms at the unification
scale) that can explain the discrepancy without contradicting  other
experimental results.


\section{Origin of FCNC in GUT Models}

In SUSY theories, the SUSY breaking terms can be the sources of
 flavor violations.
In general, it is easy to include  sources of flavor
violation  by hand and the discrepancy between the unitarity
condition and the experimental measurement can be fitted since the SUSY
breaking masses with flavor indices are parameters in the model.
However, if these parameters are completely general, too much FCNCs
are induced.
Therefore, as a minimal assumption of the SUSY breaking,
  universality is often considered,
which means that all the SUSY breaking scalar masses are universal
to be $m_0^2$, and the scalar trilinear couplings are proportional
to Yukawa couplings (the coefficient is universal to be $A_0$) at
unification scale. In this case, the angles in the unitarity
triangle are not changed. Since the measurements of the $V_{ub}$ are
obtained from tree-level processes, the SUSY particles  do not
contribute to its determination. Therefore, we need to include
flavor violating sources at the unification scale to explain the
discrepancy.

In GUT models, the flavor violating sources in SUSY breaking
parameters can be induced from neutrino Dirac Yukawa couplings,
$Y_\nu \,\bar {\bf 5} N^c H_{{\bf 5}}$ \cite{Borzumati:1986qx,Barbieri:1994pv}.
 Since the left-handed lepton doublet, $L$, and the right-handed down-type
quarks, $D^c$, are unified in $\bar{\bf 5}$ multiplet, the SUSY
breaking mass matrix for $D^c$ is corrected by colored-Higgs and
right-handed neutrino loops. As a result, the flavor violation in
the quark sector can be generated from the neutrino coupling. The
contribution is naively proportional to $Y_\nu Y_\nu^\dagger$. Our
purpose in this section is to search for a solution of the
$V_{ub}$-$\sin2\beta$ discrepancy using the neutrino mixings in the
context of a GUT scenario.

We will work in a basis where the right-handed neutrino Majorana
mass matrix, $M_R$, and charged-lepton Yukawa matrix, $Y_e$, are
diagonal,
\begin{equation}
M_R = {\rm diag}\, (M_1,M_2,M_3).
\end{equation}
The neutrino Dirac Yukawa coupling matrix is written as
\begin{equation}
Y_\nu = V_L^e Y_\nu^{\rm diag} V_R^{e\dagger},
\label{neutrino-Dirac}
\end{equation}
where $V_{L,R}^e$ are diagonalizing unitary matrices.
We note that $V_L^e$ corresponds to the (conjugate of)
MNSP neutrino mixing matrix, $U_{\rm MNSP}$,
in type I seesaw, $m_\nu^{\rm light} = Y_\nu M_R^{-1} Y_\nu^{\rm T}
\langle H_u^0 \rangle^2$
 \cite{Minkowski:1977sc},
up to a diagonal phase matrix when $V_R^e = {\bf 1}$ (identity matrix),
which we will assume for simplicity.
Through RGE, the off-diagonal elements of
the SUSY breaking mass matrix for the left-handed lepton doublet
gets the following correction
\begin{equation}
\delta M_{\tilde L}^2{}_{ij} \simeq -\frac{1}{8\pi^2} (3 m_0^2 + A_0^2)\,
\sum_k (Y_\nu)_{ik} (Y_\nu^*)_{jk} \ln \frac{M_{*}}{M_k},
\end{equation}
where $M_*$ is a cutoff scale and
the SUSY breaking parameters are universal.
Neglecting the threshold of the GUT and the Majorana mass scales,
we can write down the boundary conditions as
\begin{equation}
M_{\bar{\bf 5}}^2 =
M_{\tilde D^c}^2 = M_{\tilde L}^2 = m_0^2
\left(
{\bf 1} - \kappa\, V_L^e
\left(
\begin{array}{ccc}
k_1 & & \\
& k_2 & \\
&& 1
\end{array}
\right) V_L^{e\dagger}
\right),
\label{boundary-5}
\end{equation}
where $\kappa
\simeq (Y_\nu^{\rm diag})_{33}^2
(3+A_0^2/m_0^2)/8\pi^2 \ln M_*/M_{\rm GUT}$, and
$k_2 \simeq \sqrt{\Delta m_{\rm sol}^2/\Delta m_{\rm atm}^2}
M_2/M_3$.
%
We parameterize the unitary matrix $V_L^e$ as
\begin{equation}
 V_L^e =
\left(\begin{array}{ccc}
e^{i (\alpha_1 - \delta)} & & \\
& e^{i \alpha_2} & \\
& & 1
\end{array}
\right)
\left(
\begin{array}{ccc}
c^e_{12} c^e_{13} & s^e_{12} c^e_{13} & s^e_{13}e^{i \delta} \\
-s^e_{12} c^e_{23} - c^e_{12}s^e_{23} s^e_{13}e^{-i \delta} &
c^e_{12} c^e_{23} - s^e_{12} s^e_{13} s^e_{23} e^{-i \delta} &
c^e_{13} s^e_{23} \\
s^e_{12} s^e_{23}- c^e_{12} c^e_{23} s^e_{13} e^{-i \delta} &
-c^e_{12} s^e_{23} -s^e_{12} s^e_{13} c^e_{23} e^{-i \delta} &
c^e_{13} c^e_{23}
\end{array}
\right),
\label{MNSP}
\end{equation}
where $s_{ij}^e$ and $c_{ij}^e$ are sin and cos of mixing angles
$\theta_{ij}$. In the limit $k_{1,2} \to 0$, $\alpha_1$ and
$\alpha_2$ are the phase of the 13 and the 23 element of $M_{\bar
{\bf 5}}^2$.
Since we are assuming that $V_R^e = \bf 1$, $\theta_{12}$ and
$\theta_{23}$ correspond to solar and atmospheric neutrino mixings,
respectively, which are large. On the other hand, $s_{13}^e$ is
bounded by CHOOZ experiments, $s_{13}^e \alt 0.2$
\cite{Apollonio:1999ae}. The SUSY breaking mass for $\bf 10$
multiplet $(Q,U^c,E^c)$ is also corrected by the (colored-)Higgsino loop, but
it arises from CKM mixings and the effect is small. So, we assume
that the boundary condition at the GUT scale for $\bf 10$ multiplet
is
\begin{equation}
M_{\bf 10}^2 = M_{\tilde Q}^2 = M_{\tilde U^c}^2 = M_{\tilde E^c}^2
= m_0^2 \,{\bf 1},
\label{boundary-10}
\end{equation}
neglecting the small contribution  arising from the CKM mixings.
The boundary conditions, Eqs.(\ref{boundary-5},\ref{boundary-10}),
are typical assumptions in the case of SU(5) GUT.
The Yukawa coupling matrices for up- and down-type quarks and
charged-leptons are given as
%
%
\begin{eqnarray}
Y_u &=& V_{qeL} V_{\rm CKM}^{\rm T} Y_u^{\rm diag} P_u V_{uR},
\label{Yukawa-u} \\
Y_d &=& V_{qeL} Y_d^{\rm diag} P_d V_{qeR}^{\rm T},
\label{Yukawa-d} \\
Y_e &=& Y_e^{\rm diag} P_e,
\label{Yukawa-e}
\end{eqnarray}
where $Y_{u,d,e}^{\rm diag}$ are real
(positive) diagonal matrices and $P_{u,d,e}$ are diagonal phase
matrices.
In a minimal SU(5) GUT, in which only $H_{\bf 5}$ and
$H_{\bar {\bf 5}}$ couple to matter fields,
we have $V_{uR} = V_{\rm CKM} V_{qeL}^{\rm T}$,
$V_{qeL} = V_{qeR} = {\bf 1}$,
and $Y_d^{\rm diag} = Y_e^{\rm diag}$.
We may consider non-minimal SU(5), but
we assume $V_{uR}, V_{qeL}, V_{qeR} \simeq \bf 1$.
Otherwise, the simple relations between the  flavor violations
in the  quark and lepton sectors which we see below
will be lost.

All matter fields are unified in the spinor representation $\bf 16$
in
 SO(10) models. Since the right-handed neutrino is also unified to other
matter fields, the
 neutrino Dirac Yukawa coupling does not
have large mixings ({\it i.e.} $V_{L}^e \simeq {\bf 1}$) in a simple
fit of the Yukawa couplings.
In this case, the proper neutrino masses with large mixings
can be generated by the type II seesaw mechanism \cite{Schechter:1980gr},
in which the
interaction term  $\frac12 f LL \Delta_L$ (where $\Delta_L$ is a SU(2)$_L$ triplet)
 induces the light neutrino masses,
$m_\nu^{\rm light} = f \langle \Delta_L^0 \rangle$. Due to the
unification under SO(10), the left-handed Majorana coupling, $f$, is
unified to the other matter fields, and the off-diagonal terms in
the sparticle masses are induced by loop effect which are
proportional to $f f^\dagger$. Neglecting the GUT scale threshold,
we can write the boundary condition in SO(10) as
\begin{equation}
M_{{\bf 16}}^2 =
m_0^2
\left(
{\bf 1} - \kappa_{16} \, U
\left(
\begin{array}{ccc}
k_1 & & \\
& k_2 & \\
&& 1
\end{array}
\right) U^\dagger
\right),
\label{boundary-16}
\end{equation}
where $\kappa_{16} \simeq 15/4\,(f_{33}^{\rm diag})^2
(3+A_0^2/m_0^2)/8\pi^2 \ln M_*/M_{\rm GUT}$,
and $k_2 \simeq \Delta m_{\rm sol}^2/\Delta m_{\rm atm}^2$
in this case.
Note that the parameters $\kappa_{16}$, $k_{1,2}$ are of course
different from those given in Eq.(\ref{boundary-5})
using the set-up for type I seesaw,
but we use the same notation to make the description  simple.
The unitary matrix $U$ is the (conjugate of) MNSP neutrino mixing
matrix up to a diagonal phase matrix, which is parameterized in the
same way as Eq.(\ref{MNSP}). The Yukawa couplings are also given as
Eq.(\ref{Yukawa-u},\ref{Yukawa-d},\ref{Yukawa-e}). If we do not
employ $\bf 120$ Higgs fields, the Yukawa matrices are symmetric, and
thus, $V_{uR} = V_{\rm CKM} V_{qeL}^{\rm T}$, $V_{qeR}= V_{qeL}$. The
unitary matrix $V_{qeL}$ is expected to be close to $\bf 1$ if there
is no huge fine-tuning in the fermion mass fits.

Now let us consider the necessary conditions to solve the
discrepancy between the unitarity condition and experimental
measurements using the boundary conditions,
Eqs.(\ref{boundary-5},\ref{boundary-10}) in SU(5) and
Eq.(\ref{boundary-16}) in SO(10) respectively.
Due to the smallness of $k_{1,2}$ and $s_{13}^e$, the 12 and 13
elements of $M_{\bar {\bf 5}}^2$ and $M_{\bf 16}^2$ are expected to
be smaller than the 23 element. In the limit of $k_{1,2}, s_{13}^e
\to 0$, only the 23 element is non-zero and then $\sin2\beta$ can
not be modified. In this limit, the $B_s$-$\bar B_s$ mass difference
$\Delta m_s$ is modified, and therefore $|V_{td}/V_{ts}|$ prediction
is changed, and one of the sides of the unitarity triangle is modified.
However, if $|V_{td}/V_{ts}|$ is modified to obtain larger $|V_{ub}|$
from the unitarity relation, CP violation in $K$-$\bar K$ mixing, $\epsilon_K$,
will not fit well with the experimental range.
Thus, we need to modify $\sin 2 \beta$ itself employing the finite
contributions of the 13 element in SUSY breaking mass matrix. If at
least one of $k_2$ and $s_{13}^e$ is non-zero, the 13 element is
generated. Then the 12 element of SUSY breaking mass matrix
is also generated in general. However,
if the 12 element is generated,
 due to the quark-lepton unification,
$\mu \to e\gamma$ process is enhanced through the chargino diagram.
In order to satisfy the experimental bound for the branching ratio of $\mu \to e\gamma$
process, Br($\mu\to e\gamma) < 1.2 \times 10^{-11}$
\cite{Brooks:1999pu}, the 12 element needs to be small to suppress
the chargino contribution. It can happen by a cancellation when $k_1
\ll k_2$, $s_{13}^e \sim k_2 \sin2\theta_{12}/2$ and $\delta \simeq
\pi$.

In order to obtain a sizable effect in $\sin2\beta$, $\kappa$ needs
to be large enough. However, if $\kappa$ is too large, $\tau \to
\mu\gamma$ process may exceed the experimental bound,
Br($\tau\to\mu\gamma)< 4.5 \times 10^{-8}$
\cite{Aubert:2005ye} for large  $\theta_{23}$ mixing. A large value
of $m_0 \sim 1$ TeV can avoid the excess of $\tau\to\mu\gamma$ due
to the following reason.
The off-diagonal elements of SUSY breaking mass matrices are
proportional to $m_0^2$, whereas the diagonal element of squark
matrix at the weak scale is insensitive to $m_0$ when $m_0 \alt 500$
GeV for gluino mass $\sim 1$ TeV, since RGE contribution from the gluino
is large. Consequently, the flavor violations from SUSY
contributions in the quark sector become maximal for $m_0 \sim 1$
TeV.
On the other hand, the diagonal elements of slepton mass matrix are
sensitive to $m_0$, and thus, lepton flavor violating processes are
suppressed  for large $m_0$.

Now let us study different results that can be  produced by type I
seesaw for SU(5) boundary condition,
Eq.(\ref{boundary-5},\ref{boundary-10}), and type II seesaw for
SO(10) boundary conditions, Eq.(\ref{boundary-16}).
If the boundary condition is flavor universal, which means
$\kappa=0$, the chargino contribution is the dominant SUSY
contribution to the $B_{d,s}$-$\bar B_{d,s}$ mixing amplitude,
$M_{12}(B_{d,s})$.
Using a  general parameter space for the soft SUSY breaking terms,
the gluino box diagram dominates the SUSY contribution.
The gluino contribution (divided by the SM contribution) can be
written naively in the following mass insertion form
\begin{equation}
\frac{M_{12}^{\tilde g}}{M_{12}^{\rm SM}}
\simeq
a\, [(\delta_{LL}^d)_{3i}^2+ (\delta_{RR}^d)_{3i}^2]
- b \, (\delta_{LL}^d)_{3i} (\delta_{RR}^d)_{3i},
\label{gluino-contribution}
\end{equation}
where $i=1$ for $B_d$ mixing, $i=2$ for $B_s$ mixing,
 $a$ and $b$ depend on squark and gluino masses, and
$\delta_{LL,RR}^d = (M^2_{\tilde d})_{LL,RR}/\tilde m^2$ ($\tilde m$
is an averaged squark mass). The matrix $M^2_{\tilde d}$ is a
down-type squark mass matrix $ (\tilde Q, \tilde D^{c\dagger})
M^2_{\tilde d} (\tilde Q^\dagger, \tilde D^c)^{\rm T} $
at weak scale in the basis
where down-type quark mass matrix is real (positive) diagonal.
When the squark and the gluino masses are less than 1 TeV, $a \sim
O(1)$ and $b \sim O(100)$. We also have contributions from
$\delta_{LR}^d$, but we neglect them  since they are suppressed by
$(m_b/m_{\rm SUSY})^2$.

Using the SU(5) boundary condition, Eq.(\ref{boundary-5},\ref{boundary-10}),
we find that $\delta_{RR}^d$ can be large due to large neutrino mixings,
 but $\delta_{LL}^d$ can not be very large
since it is induced from CKM related mixing. On the other hand,
using the SO(10) boundary conditions we find that both
$\delta_{LL}^d$ and $\delta_{RR}^d$ can be large. As a result, the
SO(10) boundary condition can generate larger SUSY contribution in
 $B$-$\bar B$ mixings rather than the SU(5) boundary condition
for the same parameters for $\kappa$, $k_{1,2}$ and mixing angles
$\theta_{ij}$ since $b \gg a$ in Eq.(\ref{gluino-contribution}).

The $B$-$\bar B$ mixing amplitude is given as
\begin{equation}
M_{12}(B)^{\rm full} = M_{12}(B)^{\rm SM} + M_{12}(B)^{\rm SUSY}.
\end{equation}
The angle $2\beta$ for unitarity triangle
is the argument of $M_{12}(B_d)^{\rm SM}$
in the particle data group phase notation of CKM matrix.
The observed angle $2\beta^{\rm eff}$
by $B_d \to J/\psi K$ decay
is the argument of $M_{12}(B_d)^{\rm full}$.
The SUSY contribution of
$B_d$-$\bar B_d$ mixing is naively proportional to
$(s_{13}^e){}^2$.
As was noted, it is necessary that $s_{13}^e \sim k_2 \sin
\theta_{12}/2$ to suppress $\mu\to e\gamma$ process. In the case of
type I seesaw in SU(5) model, $k_2$ is a free parameter, and thus,
$s_{13}^e$ is free. On the other hand, using type II seesaw in the
SO(10) model, $k_2$ is almost determined by the ratio of mass
squared differences for solar and atmospheric neutrino oscillations.
As a result, the suppression of $\mu\to e\gamma$ process indicates
$s_{13}^e \sim 0.02$.
Due to the smallness of $s_{13}^e$, the modification of $\sin2\beta$
will be small for type II seesaw SO(10) boundary condition. For type
I seesaw SU(5) boundary condition, on the other hand, the
modification can be larger than the SO(10) case since $s_{13}^e$ can
be sizable $s_{13}^e \sim 0.1$, though the modification is smaller
if we use the mixing angle $s_{13}^e \sim 0.02$.

The absolute value   of $M_{12}(B_{d(s)})^{\rm SUSY}$
is almost determined by $\kappa$, $s_{13}^e$ ($s_{23}^e$),
and squark, gluino masses,
whereas,
the argument of $M_{12}(B)^{\rm SUSY}$
can be anything
since $\alpha_{1,2}$ in Eq.(\ref{MNSP})
and phases $P_{d,u}$ in Yukawa matrices in
Eq.(\ref{Yukawa-u},\ref{Yukawa-d})
are all free.
The minimal value of $\sin2\beta^{\rm eff}$
for a given size of SUSY contribution
can be obtained
by choosing the phases.
The branching ratios of lepton flavor violating processes
do not depend on the phases
since the branching ratios are proportional to
the squared absolute values of the amplitudes of
SUSY contributions.

\begin{figure}[tb]
 \center
 \includegraphics[viewport = 20 20 280 225,width=8cm]{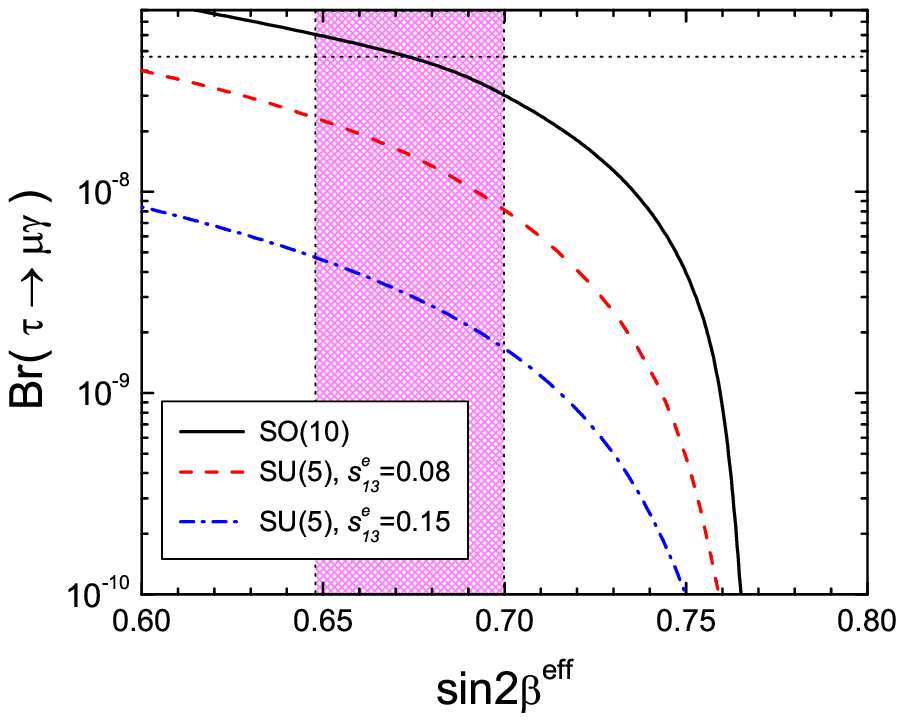}
 \includegraphics[viewport = 20 20 280 225,width=8cm]{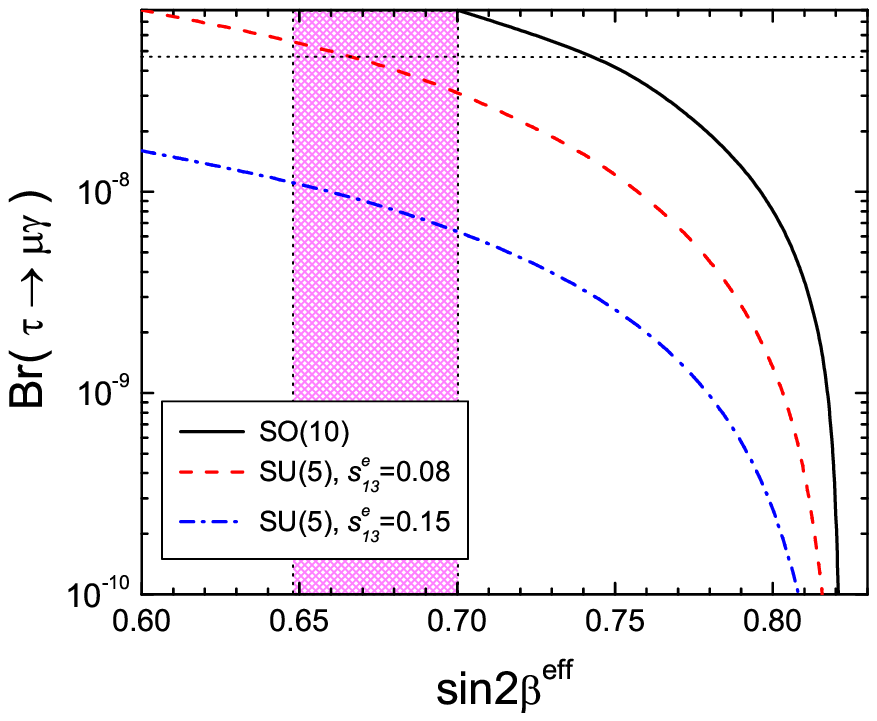}
 \caption{
  Relation of the minimal value of $\sin 2\beta^{\rm eff}$
  and branching ratio of $\tau\to\mu\gamma$ decay
  for $|V_{ub}| = 0.0041$ (left), $0.0045$ (right).
  The dotted line stands for the experimental bounds.
 }
\end{figure}

We plot the relation between the minimal value of $\sin 2\beta^{\rm
eff}$ and the branching ratio of $\tau\to\mu\gamma$ decay in Figure
2. In the plot, we use $m_0 = 1.2$ TeV, $m_{1/2} = 300$ GeV for
gaugino mass at GUT scale, and $\tan\beta_H =10$ for a ratio of
vacuum expectation values of up- and down-type Higgs fields. In the
SU(5) case, one can tune $k_2$ for a given $s_{13}^e$ to suppress
Br($\mu \to e\gamma)$ to be less than $10^{-11}$. With type II
seesaw in the SO(10) model,
 $k_2$ is fixed by observation of neutrino oscillation and the
suppression of Br($\mu\to e\gamma$) implies $s_{13}^e \sim 0.02$.
Since there are also neutralino contributions for both left- and
right-handed muon decay, one can not cancel Br($\mu \to e\gamma)$
completely. In the range where $\sin 2\beta^{\rm eff}$ is in
agreement with experiment, we obtain Br($\mu \to e\gamma) \agt
10^{-12}-10^{-13}$. In the plot, we show the cases $s_{13}^e =
0.08$, and $0.15$ for SU(5), and $s_{13}^e = 0.022$ for SO(10). As
was expected, one can read off that the modification of
$\sin2\beta^{\rm eff}$ can become large for larger $s_{13}^e$ for
the same value of Br($\tau\to \mu\gamma$).

\begin{figure}[tb]
 \center
 \includegraphics[viewport = 20 20 280 225,width=8.5cm]{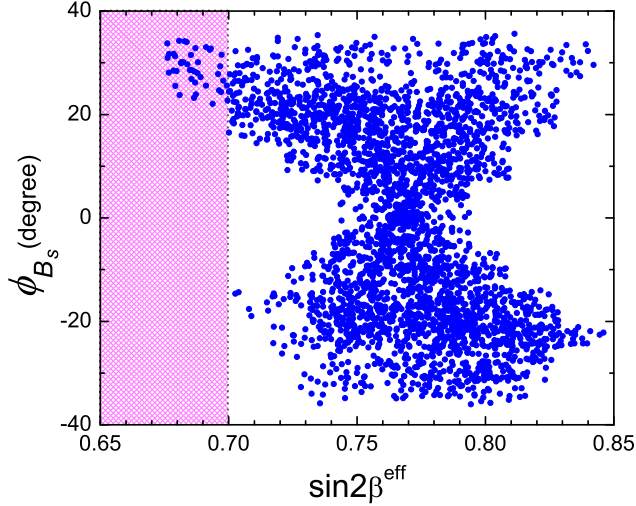}
 \caption{Random dot plot for $\sin2\beta^{\rm eff}$ and $\phi_{B_s}$
 for $|V_{ub}| = 0.0041$
 using the SO(10) boundary condition. The shaded area shows the 1$\sigma$
 range of $\sin2\beta$ for the world average \cite{Barberio:2006bi}.
 }
\end{figure}

So far we have obtained the  parameter region which satisfies the observed
$\sin2\beta^{\rm eff}$ and the inclusive data of $|V_{ub}|$, as well
as the bounds on the leptonic flavor violating decay in a grand
unified scenario. We will next study the consequence of the fit in
the phase of $B_s$-$\bar B_s$ mixing. Although the measurements of
mass differences, $\Delta M_{d,s} \equiv 2 |M_{12}(B_{d,s})^{\rm
full}|$, are in good agreement with SM prediction, there can still
be room for SUSY contribution due to the following two reasons. One
reason is that there exists large uncertainties $\sim 20$\% in the lattice
calculation for the decay constants $f_{B_{d,s}}$ and the bag
parameters $B_{B_{d,s}}$.
The SM prediction of the ratio $\Delta M_s/\Delta M_d$ can be more
accurate since the ratio for  $f_{B}$ and $B_B$ have less
uncertainty. The other reason is that we still have an ambiguity for
the phase of $B_s$-$\bar B_s$ mixing. For instance, one can satisfy
$|M_{12}(B_s)^{\rm full}| = |M_{12}(B_s)^{\rm SM}|$ by choosing a
phase of $M_{12}(B_s)^{\rm SUSY}$ as long as $|M_{12}(B_s)^{\rm
SUSY}|< 2|M_{12}(B_s)^{\rm SM}|$. As we have noted, the phase of
$M_{12}(B_s)^{\rm SUSY}$ is completely free for the boundary
conditions in GUT models.
The SM prediction of the phase of
$B_s$-$\bar B_s$ oscillation ($2\beta_s = -{\rm arg}\,M_{12}(B_s)^{\rm SM}$)
is about $2^{\rm o}$
($\sin 2\beta_s \simeq 0.04$).
If the SUSY contribution modifies  $\sin 2\beta^{\rm eff}$ and
resolves the tension of the $\sin2\beta$-$V_{ub}$ unitarity
relation, the phase of $B_s$-$\bar B_s$ ($2\beta_s^{\rm eff} = -
{\rm arg}\,M_{12}(B_s)^{\rm full}$) must be different from the SM
prediction.

It is convenient to define the deviation from SM prediction
by two real parameters $C_{B_s}$ and $\phi_{B_s}$
as
\begin{equation}
C_{B_s} e^{2 i \phi_{B_s}} \equiv \frac{M_{12}(B_s)^{\rm full}}{M_{12}(B_s)^{\rm SM}}\,,
\label{Cbs-phibs}
\end{equation}
If there is no SUSY contribution, $C_{B_s} = 1$ and $\phi_{B_s}= 0$.
In this notation, $\beta_s^{\rm eff} = \beta_s - \phi_{B_s}$. We are
interested in the value of $\phi_{B_s}$ in the range where
$\sin2\beta_{\rm eff}$ is in agreement with the experimental data
for inclusive $V_{ub}$. In the case of SU(5) boundary condition, the
value is $\phi_{B_s} \sim \pm 3^{\rm o}$, namely $\sin 2\beta_s^{\rm
eff} \sim +0.14, -0.07$. On the other hand, in the case of SO(10)
boundary condition, $\phi_{B_s}$ can be sizable. We show
$\phi_{B_s}$ and $\sin2\beta_{\rm eff}$ in SO(10) type II seesaw by
random dot plot in Figure 3. We choose the same parameters as in the
plot for Figure 2. Constraint arising from the Br($\tau
\to\mu\gamma) < 4.5 \times 10^{-8}$ rules out larger values of
$|\phi_{B_s}| \agt 35^{\rm o}$. We also filter out the points by the
constraint from $K$-$\bar K$ mixing, and $\Delta M_s/\Delta M_d$.
Especially, points at the left-bottom corner in the plot are ruled
out by  $K$-$\bar K$ mixing data. The reason for the allowed
left-top corner is that the  SUSY contribution becomes
 $\epsilon_K^{\rm SUSY} \sim  - 2 \epsilon_K^{\rm SM}$.
{}From the figure, we find the  preferred value to be $\phi_{B_s} \sim
 +(20-30)^{\rm o}$, which corresponds to
$\sin2\beta_s^{\rm eff} \sim - (0.6 -0.9)$.
The recent measurement by D\O~collaboration \cite{D0} shows
a large negative central value for $\sin 2\beta_s^{\rm eff} = -0.71$
with a sizable error
and if this result holds then, definitely,
the SO(10) model will be preferred.

We summarize what we have accomplished in this section. We have
considered the parameter space where we can resolve  the discrepancy
between the unitarity prediction and the experimental measurement
for the boundary conditions for SU(5) with type I seesaw and for
SO(10) with type II seesaw. To obtain the solution we need large  $m_0
\sim 1$ TeV to suppress $\tau \to \mu\gamma$ decay.
In order to suppress $\mu\to e\gamma$, we need to adjust the 13
mixings in the Dirac Yukawa couplings for type I seesaw and the Majorana
couplings for type II seesaw. Consequently, the neutrino oscillation
parameter is constrained as follows. The CP phase in neutrino
oscillation $\delta_{\rm MNSP}$ needs to be close to $\pi$. In type
II seesaw case, the 13 neutrino mixing should be around 0.02, while
in type I seesaw case, the 13 mixing is free.
The phase of $B_s$-$\bar B_s$ is not same as that can be obtained in
the SM. The deviation is small for the SU(5) case, and large for the
SO(10) case.

The crucial assumptions we made to obtain above results are that
$V_{R}^e \simeq {\bf 1}$ in the Dirac Yukawa coupling,
Eq.(\ref{neutrino-Dirac}), in type I seesaw for SU(5), and  the
triplet part dominates the light neutrino mass rather than the type
I seesaw part for type II seesaw for SO(10). If we do not assume
them, $V_L^e$ and $U$ in the boundary conditions do not take part in
the construction of the neutrino mixing matrix. Even in this case,
we may need the relation, $s_{13}^e \sim k_2 \sin2\theta_{12}/2$,
and $\delta \simeq \pi$ to suppress $\mu \to e\gamma$ decay, but the
mixing angles and the phase are not necessarily related to the
observed neutrino mixings. Moreover, since $s_{23}^e$ is not
necessarily large, $\tau \to \mu\gamma$ can be suppressed by
choosing a small $s_{23}^e$ even for smaller $m_0 \alt 500$ GeV. If
this is the case, depending on
the $s_{23}^e$, the deviation of the $B_s$-$\bar B_s$ phase
$\phi_{B_s}$ will not be large compared to the above. The possible feature of this case is that
Br($\tau\to e\gamma$) may be comparable to Br($\tau \to\mu\gamma$)
since $s_{13}^e$ needs to be around 0.1 to modify $\sin2\beta^{\rm
eff}$ while  we get Br($\tau\to e\gamma$) $\ll$ Br($\tau
\to\mu\gamma$) in a usual scenario.

\section{Origin of FCNC Using Non-proportional $A$-term}

In the previous section, we have considered the scenario where the
neutrino mixings are the origin of flavor violation at low energy,
assuming the universality for SUSY breaking scalar masses and
trilinear scalar couplings at unification scale. The SUSY breaking
scalar mass and trilinear scalar couplings are given as
\begin{equation}
m_{a\bar a}^2 = m_{3/2}^2 + V_0 - \sum_{M,\bar N} F^{M} \bar F^{\bar N}
\partial_{M} \partial_{\bar N} \ln K_{a\bar a},
\end{equation}
\begin{equation}
A_{abc} = \sum_M F^M \left[(K_M - \partial_M \ln (K_{a\bar a} K_{b\bar b} K_{c\bar c}))
           Y_{abc}
         + \partial_M Y_{abc}\right],
\end{equation}
where $K$ is the K\"ahler potential and $F^M$ is the $F$-term of
moduli field $M$. The flavor invisibleness of K\"ahler potential is
necessary for the universality of SUSY breaking parameters. However,
the invisibleness is not a sufficient condition to make the $A$-term
flavor universal. Actually, when the Yukawa coupling depends on moduli
and the $F$-term of the moduli is non-zero, the $A$-terms are no
longer proportional to Yukawa coupling, even if the K\"ahler potential
is flavor invisible. In the intersecting D-brane models
\cite{Blumenhagen:2005mu}, the K\"ahler potential can be flavor
invisible, but Yukawa couplings depend on $U$-moduli (shape moduli),
and they can generate flavor violation in the SUSY breaking scalar
mass matrices at low energy \cite{Dutta:2006bp}. In this section, we
will consider non-proportional $A$-terms to resolve the
$\sin2\beta$-$V_{ub}$ tension.

In the intersecting D-brane models, the SUSY breaking scalar masses
are flavor universal when the generations are simply replicated at
the intersection of the branes. The Yukawa couplings are induced
by the three-point open string scattering, and they are determined
by the triangle area formed by branes. As we have noted, the
couplings depends on $U$-moduli. The Yukawa coupling matrices
are given in a factorized form \cite{Cremades:2003qj},
\begin{equation}
Y_{ij}^0 = x_i^L (U_1) x_j^R (U_2),
\label{Yukawa-rank1}
\end{equation}
if the left- and right-handed matter fields  are replicated on different
tori. The couplings are the function of Jacobi theta function
$\vartheta \left[ \begin{array}{c} \delta \\ \phi \end{array}\right]
(t)$. The coupling $x_i$ is given as
\begin{equation}
x_1 : x_2 : x_3 =
\vartheta \left[ \begin{array}{c} \varepsilon+\frac13 \\ 0 \end{array}\right] (t)
:
\vartheta \left[ \begin{array}{c} \varepsilon-\frac13 \\ 0 \end{array}\right] (t)
:
\vartheta \left[ \begin{array}{c} \varepsilon \\ 0 \end{array}\right] (t),
\label{x1x2x3}
\end{equation}
where $\varepsilon$ stands for a brane shift parameter,
and the Wilson line phase $\phi$ is chosen to be zero for simplicity.
The variable $t$ is proportional to  $U$-moduli.
Since the Yukawa matrices are rank 1, only third generation becomes
massive after the electroweak phase transition. We, therefore, need
to introduce corrections to  Yukawa couplings, which can arise due
to higher dimensional operators. If the correction $\delta Y$ is
given as $\delta Y = {\rm diag}\,(0,0,\epsilon)$ in the basis where
the light neutrino Majorana mass is diagonal, $U_{e3}$ (the 13
mixing in neutrino oscillation) is exactly zero. In the quark
sector, the correction $\delta Y = {\rm diag}\,(0,0,\epsilon)$ leads
to the relation $m_s/m_b \sim V_{cb}$. The first generation fermions
are still massless with the above  correction to the rank 1 Yukawa
matrix and we need more corrections. Thus, $U_{e3}$ is small and
relates to  the size of the  Cabibbo angle in the quark-lepton
unification picture. Furthermore, $\tan \theta_{\rm sol} \alt 1$ for
the solar mixing angle, and $\sin^2 2\theta_{\rm atm} \sim 1$ for
the atmospheric neutrino mixing can be satisfied naturally in the
perturbative region of the Yukawa coupling. The details can be found
in the Refs.~\cite{Dutta:2005bb,Dutta:2006bp}.

If the $F$-terms of  $U$-moduli, $F^U$, are zero
and the  K\"ahler potential is flavor invisible,
the $A$-term is proportional to the Yukawa couplings.
We consider the case where the K\"ahler potential is flavor invisible
whereas $F^U \neq 0$.
The rank 1 matrix $Y^0$ is given as in Eq.(23).
For simplicity, we consider that the matrix is symmetric:
$x^{L,R}_i = x_i(U)$.
The diagonalization unitary matrix of $Y^0$ is
\begin{equation}
U^0 = \left( \begin{array}{ccc}
                 \cos \theta_s & - \sin\theta_s & 0 \\
                 \cos \theta_a \sin\theta_s & \cos\theta_a \cos\theta_s & -\sin\theta_a \\
                 \sin \theta_a \sin\theta_s & \sin\theta_a \cos\theta_s & \cos\theta_a
               \end{array}
        \right),
\end{equation}
where $\tan\theta_s = x_1/x_2$ and $\tan\theta_a =
\sqrt{x_1^2+x_2^2}/x_3$. We work in a basis where the light neutrino
mass is diagonal assuming that the correction  for the
charged-lepton Yukawa matrix is $\delta Y_e \simeq {\rm
diag}(0,0,\epsilon)$. Then, MNSP neutrino mixing matrix is given as
\begin{equation}
U_{\rm MNSP} = V_e^* U^0,
\end{equation}
where $V_e$ is a diagonalization matrix of $U^0 (Y^0+\delta Y_e)
(U^0)^{\rm T}$. If the quark-lepton unification is realized, we
obtain $V_e \simeq V_{\rm CKM}^{\dagger}$. The solar and atmospheric neutrino
mixing is close to $\theta_{s,a}$ respectively, but solar mixing is
corrected by $(V_e)_{12}$. The 13 neutrino mixing is given as
$U_{e3} \simeq (V_e)_{12}/\sqrt2$.

The non-proportional part of $A$-term is proportional to $\partial_U Y$.
The derivative of Yukawa coupling $Y^0$ is calculated
as
\begin{equation}
U^0 (\partial_U Y^0) (U^0)^{\rm T} =
\left(
\begin{array}{ccc}
0 & 0 & y \sin\theta_a \partial_U \theta_s \\
0 & 0 & y \, \partial_U \theta_a \\
y\sin\theta_a \partial_U \theta_s & y \,\partial_U \theta_a & \partial_U y
\end{array}
\right),
\end{equation}
where $y = \sum x_i^2$. The $ij$ $(i,j=1,2)$ elements of $\partial
Y^0$ are zero, and thus, the 12 element of non-proportional the
$A$-term is suppressed to be of the  order of $\sim \lambda^2
\partial Y_{13} + \lambda^3 \partial Y_{23}$ (where $\lambda \sim
0.2$) in the basis where the charged-lepton Yukawa matrix is
diagonal. This is a general property of the almost rank 1 Yukawa
matrix, independent of any set-up.

We note that the non-proportional pieces of the $A$-term can be
considered in the non-minimal scenario of hidden sector model with
flavor symmetries. However, in such a scenario, the non-proportional
 $A$-term is hierarchical and thus, large 13 off-diagonal elements are
not expected. On the other hand, in our scenario of  $U$-moduli
origin of the $A$-terms, we can realize large 13 elements of the $A$-term.

In general, if the $A$-term has non-proportional pieces, the
off-diagonal elements of SUSY breaking scalar mass matrices are
generated via RGEs, e.g. $d M^2_{\tilde D^c}/\ln Q = 1/(4\pi^2)
A_d^{\rm T} A_d^* + \cdots$. Since our purpose is to modify $\sin
2\beta$, we need the 13 (31) element of the $A$-term for left
(right) handed squark mass matrices in the basis where Yukawa
matrices are diagonal. The RGE induced 13 element of SUSY breaking
mass matrices can be larger when $A_{33}$ (=$A_b$) is large.
Therefore, a large coefficient of $A$-term, {\it i.e.} $A_0$, is
preferable. The negative $A_0$ can generate large contribution to
$\sin2\beta$ for gluino mass $M_3>0$ since the RGE is given as $d
A_b/d\ln Q = + 8/(3\pi) \alpha_3 M_3 y_b +\cdots$, where $y_b$ is a
bottom quark Yukawa coupling.

When both $A_{13}$ and $A_{23}$ elements are introduced at boundary
condition, the 12 elements of SUSY breaking mass matrices are
generated at low energy even if $A_{12}$ is zero in the boundary
condition.
 The 12 elements of SUSY breaking
mass matrices are unwanted for $K$-$\bar K$ mixing and $\mu\to
e\gamma$. Thus, for our purpose to modify $\sin2\beta$, we need to
suppress $A_{23}$ ($A_{32}$) rather than $A_{13}$ ($A_{31}$). Such a
situation can be satisfied when $\partial_U \theta_a = 0$.
There is a solution of $\partial_U \theta_a = 0$ when $\frac16
<\varepsilon < \frac14$ where $\varepsilon$ is a brane shift
parameter in Eq.(\ref{x1x2x3}). Since $\partial_U \theta_a= 0$, the
atmospheric mixing as a function of $U$ is maximized for a given
$\varepsilon$. The predicted atmospheric mixing is $\frac89 <\sin^2
2\theta_a<1$ and $\theta_a > 45^{\rm o}$.

For the numerical studies,
we assume the following boundary condition for the non-proportional $A$-term:
\begin{equation}
A_u^{\rm np} = A_d^{\rm np} = A_e^{\rm np} =
\left(
\begin{array}{ccc}
0 & 0 & \delta \\
0 & 0 & 0  \\
\delta & 0 & 0
\end{array}
\right) A_0\,,
\label{np-A}
\end{equation}
in the basis where Yukawa matrices are diagonal.
The $A$-terms for up- and down-type quarks and charged-leptons
are given as $A_{u,d,e} = A_0 Y_{u,d,e} + A_{u,d,e}^{\rm np}$.
In general,
there is no reason that $A^{\rm np}$'s are unified,
but we assume the unification in the context of
$G_{422}$ = SU(4)$_c \times $ SU(2)$_L \times $ SU(2)$_R$ model construction
of intersecting D-brane:
The $A^{\rm np}$ may originate from $G_{422}$ invariant term,
while the corrections to the  Yukawa coupling breaks $G_{422}$.
The $A^{\rm np}$ is assumed to be symmetric  just for simplicity.

\begin{figure}[tb]
 \center
 \includegraphics[viewport = 20 20 280 225,width=8.5cm]{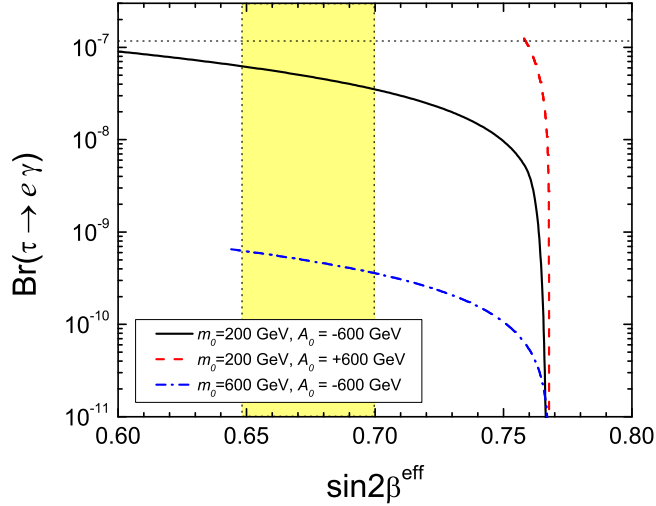}
 \caption{The $\sin2\beta^{\rm eff}$ and Br($\tau \to e\gamma$) relation
employing non-proportional $A$-term.
The dotted lines stand for the experimental bounds.
 }
\end{figure}

We plot  $\sin2\beta^{\rm eff}$ and Br($\tau\to e\gamma$) in Figure
4, varying $\delta$ in Eq.(\ref{np-A}) from 0 to $y_b$. In the plot,
we use $m_0= 200$ GeV, $m_{1/2} = 300$ GeV, $\tan\beta_H = 10$ and
$|V_{ub}| = 4.1 \times 10^{-3}$. When sgn$(y_d) = - $sgn$(y_b)$ for
Yukawa couplings of down and bottom quarks, the SUSY contribution of
$\sin2\beta^{\rm eff}$ becomes negative and satisfy the observation
from CP asymmetry of $B_d \to J/\psi K$. The positive values of
$A_0$ do not modify $\sin2\beta$ much as we have explained. The
Br($\tau\to e\gamma$) do not depend on the signature of $A_0$
because RGE running of $A_\tau$ is not large.

Since we have assumed that $A_{12}$ and $A_{23}$ are zero,
$\epsilon_K$ and $B_s$-$\bar B_s$ mixing ($\Delta M_s$ and
$\phi_{B_s}$) do not get modified from the SM values.
We do not get any enhancement in the branching ratios of $\mu\to e\gamma$ and
$\tau \to \mu\gamma$, neither.
When we introduce $A_{23}$ elements, all these
quantities can be modified. The strongest constraint comes form
$\epsilon_K$ when $\sin2\beta^{\rm eff}$ is in the experimental
range.

Interestingly, in this scenario we have  solutions for smaller
values of $m_0$.
Thus, the solution can be consistent 
with the muon $g-2$ from $e^+e^-$ data \cite{Bennett:2002jb}. The
main feature of this solution is that
 Br($\tau \to e\gamma$) is larger than Br($\tau\to \mu\gamma$).

\section{Conclusion}

The measurements of $B_{d,s}$-$\bar B_{d,s}$ mass differences,
$\Delta M_{d,s}$, and the CP asymmetry of
 $B_d \to J/\psi K$, $\sin2\beta$,
determines the unitarity triangle, and  generates
predictions for other angles, CP violation in $K$-$\bar K$ mixing,
as well as $|V_{ub}/V_{cb}|$.
The predictions are consistent with
the measurements within errors except for  $|V_{ub}/V_{cb}|$. The
recent development of $|V_{ub}|$ measurement from the inclusive $B$
decay indicates a more than 2$\sigma$ discrepancy compared to the
value obtained from the unitarity relation.
This may imply the presence of new physics.

Since the inclusive $B$ decay to measure $|V_{ub}|$ is a tree level
process, it does not include new physics.
On the other hand, $B$-$\bar B$ mixing and $\sin2\beta$ arise from
box diagrams, and thus, they may include new physics contributions,
since the new particles can propagate in the diagram and  can shift
the experimental measurements of $\Delta M_{s,d}$ and $\sin 2\beta$.

We have studied whether the SUSY contributions can modify the
unitarity relation to make the experimental measurements to fit well,
and investigated the predictions for other measurements, such as the
CP asymmetry in $B_s$ decays and branching ratios of flavor
violating lepton decay modes.
As a consequence, the modification of
the unitarity relation of $|V_{ub}|$ and $\sin2\beta$ shifts the
tension to the phase of $B_s$-$\bar B_s$ mixing or
Br($\tau \to e\gamma)/$Br($\tau \to\mu\gamma$).

We assumed that the SUSY breaking scalar masses are universal at a
cutoff scale. 
The origin of  flavor violations can
be in the Yukawa interaction and scalar trilinear couplings.
We considered  two different set-ups
to modify the unitarity relation.
In the  first set-up, the flavor violation
originates from the large neutrino mixings
in SU(5) and SO(10) grand unified models.
In the second set-up,  $A$-term couplings include
pieces which are not proportional to Yukawa couplings
in the context of intersecting D-brane models.

In GUT models, responsible for quark-lepton unification,
the neutrino Dirac couplings and the Majorana couplings can induce
 the off-diagonal elements of squark mass matrices
via the GUT scale particle loops.
In order to modify the unitarity
relation appropriately, we need large contribution in 13 elements in SUSY
breaking scalar mass matrices.
However, a large 13 neutrino mixing
is bounded by experiments, since it leads to large effects in flavor
violating lepton decay such as $\tau \to \mu\gamma$ and $\mu\to
e\gamma$.
When we suppress the branching ratio of the lepton flavor violating
decays, neutrino mixing parameters and the SUSY spectrum is
restricted.
As a result, the CP violating phase in the neutrino oscillation needs to
be close to $\pi$ when the large solar and atmospheric mixings are
responsible for the structures of neutrino Dirac Yukawa coupling in
type I seesaw and for the Majorana coupling in type II seesaw.
In type II seesaw case, the 13 neutrino mixing is  restricted to be
around 0.02.
The squarks and the sleptons need to be heavy for the
solution, and the universal scalar mass $m_0$ is about 1 TeV.
As a consequence of  resolving the $V_{ub}$-$\sin2\beta$ tension by the
SUSY contribution in the GUT models,
the phase of $B_s$-$\bar B_s$ mixing  
  becomes different from the SM prediction, $\beta_s^{\rm SM} \simeq 1^{\rm o}$.
The prediction of the phase in the SU(5) model is not very different from the SM
prediction,
$\beta_s^{\rm SU(5)} \sim \pm 3^{\rm o}$,
while
it becomes large in the SO(10) boundary condition
to be $\beta_s^{\rm SO(10)} \sim \pm (20-30)^{\rm o}$.
%
%
In the SO(10) case, the contribution of $\epsilon_K$ is also large.
In order to fit the  $\epsilon_K$ data,
the negative value of $\beta_s^{\rm eff}$ is preferred.
The D\O\ collaboration has reported the direct measurement of the
phase, and the phase $\beta_s^{\rm eff}$ can be a large negative
value, $\beta_s^{\rm eff} = (-23 \pm 16)^{\rm o}$ \cite{D0}.
The combined analysis of the CP asymmetry in the $B_s \to
J/\psi \phi$ decay as well as asymmetry in semileptonic $B_s$ decay
improves   $\beta_s^{\rm eff} = (-16 \pm 12)^{\rm o}$.
The measurements are expected to become more accurate soon and the
prediction of the models can be tested.

The $A$-term can include a non-minimal term which is not
proportional to the Yukawa coupling if the Yukawa coupling is the
function of moduli and the moduli acquires a non-zero $F$-component
which breaks SUSY.
In the intersecting D-brane models, it is natural
that the SUSY breaking scalar mass matrices are flavor universal
while the $A$-term can have  non-proportional terms.
In this model,
when the atmospheric neutrino mixing angle is maximized as a
function of  $U$-moduli, only the 13 and the 31 elements of the
non-proportional $A$-term can be large.
As a result, the phase of
$B_d$-$\bar B_d$ mixing can be modified to make the unitarity
relation consistent with the $|V_{ub}|$ data.
The feature of this
model is that  Br($\tau \to e\gamma$) is larger than
Br($\tau \to\mu\gamma$).
The current bounds of the branching ratios are
Br($\tau \to\mu\gamma) < 4.5 \times 10^{-8}$
and Br($\tau \to e\gamma) <1.2 \times 10^{-7}$ \cite{Aubert:2005ye}.

Finally, we comment on the $\sin2\beta^{\rm eff}$ from
$b\to s$ penguin modes such as $B_d \to \phi K$ decay
in our solutions to resolve the $V_{ub}$-$\sin2\beta$ tension.
In GUT models, since the squarks are heavy $\sim 1$ TeV
in the solutions,
the contribution of the penguin diagrams are small.
Thus, $\sin2\beta^{\rm eff}_{B_d \to J/\psi K} \sim \sin2\beta^{\rm
eff}_{B_d \to \phi K}$.
In our solutions using the non-proportional $A$-terms, the  $b\to s$
penguin effects can become larger depending on the model parameters
since the squarks are light.
However, when 23 and 32 elements of the non-proportional $A$-term
are zero as we have assumed in Eq.(\ref{np-A}), the $b\to s$ penguin
effects are not significant.

\section*{Acknowledgments}
We thank Z. Ligeti and B. Casey for useful discussions.


\begin{thebibliography}{99}
%
%
%
%
%
%
%
%


\bibitem{Abulencia:2006ze}
  V.~M.~Abazov {\it et al.}  [D0 Collaboration],
  Phys.\ Rev.\ Lett.\  {\bf 97}, 021802 (2006)
  [hep-ex/0603029];
%
  A.~Abulencia {\it et al.} [CDF Collaboration],
  Phys.\ Rev.\ Lett.\  {\bf 97}, 062003 (2006)
  [hep-ex/0606027];
%
  hep-ex/0609040.


\bibitem{Kobayashi:1973fv}
  M.~Kobayashi and T.~Maskawa,
  Prog.\ Theor.\ Phys.\  {\bf 49}, 652 (1973).


\bibitem{Bigi:1981qs}
  I.~I.~Y.~Bigi and A.~I.~Sanda,
  Nucl.\ Phys.\ B {\bf 193}, 85 (1981).


\bibitem{Yao:2006px}
  W.~M.~Yao {\it et al.}  [Particle Data Group],
  J.\ Phys.\ G {\bf 33}, 1 (2006).


\bibitem{Bona:2006sa}
  M.~Bona {\it et al.}  [UTfit Collaboration],
  Phys.\ Rev.\ Lett.\  {\bf 97}, 151803 (2006)
  [hep-ph/0605213];
%
  JHEP {\bf 0610}, 081 (2006)
  [hep-ph/0606167];
%
  \url{http://utfit.roma1.infn.it/}.


\bibitem{Charles:2004jd}
  J.~Charles {\it et al.}  [CKMfitter Group],
  Eur.\ Phys.\ J.\ C {\bf 41}, 1 (2005)
  [hep-ph/0406184];
  \url{http://ckmfitter.in2p3.fr/}.



\bibitem{Gabbiani:1988rb}
  F.~Gabbiani and A.~Masiero,
  Nucl.\ Phys.\ B {\bf 322}, 235 (1989);
%
  J.~S.~Hagelin, S.~Kelley and T.~Tanaka,
  Nucl.\ Phys.\ B {\bf 415}, 293 (1994);
%
  F.~Gabbiani, E.~Gabrielli, A.~Masiero and L.~Silvestrini,
  Nucl.\ Phys.\ B {\bf 477}, 321 (1996)
  [hep-ph/9604387].



\bibitem{Borzumati:1986qx}
  F.~Borzumati and A.~Masiero,
  Phys.\ Rev.\ Lett.\  {\bf 57}, 961 (1986);
%
  J.~Hisano, T.~Moroi, K.~Tobe, M.~Yamaguchi and T.~Yanagida,
  Phys.\ Lett.\ B {\bf 357}, 579 (1995).


\bibitem{Barbieri:1994pv}
  L.~J.~Hall, V.~A.~Kostelecky and S.~Raby,
  Nucl.\ Phys.\ B {\bf 267}, 415 (1986);
%
  R.~Barbieri and L.~J.~Hall,
  Phys.\ Lett.\ B {\bf 338}, 212 (1994)
  [hep-ph/9408406];
%
  J.~Hisano, T.~Moroi, K.~Tobe and M.~Yamaguchi,
  Phys.\ Lett.\ B {\bf 391}, 341 (1997)
  [hep-ph/9605296].




\bibitem{Berkooz:1996km}
  M.~Berkooz, M.~R.~Douglas and R.~G.~Leigh,
  Nucl.\ Phys.\ B {\bf 480}, 265 (1996)
  [hep-th/9606139].
%
  H.~Arfaei and M.~M.~Sheikh Jabbari,
  Phys.\ Lett.\ B {\bf 394}, 288 (1997)
  [hep-th/9608167];
%
  R.~Blumenhagen, L.~Goerlich, B.~Kors and D.~Lust,
  JHEP {\bf 0010}, 006 (2000)
  [hep-th/0007024];
%
  R.~Blumenhagen, B.~Kors and D.~Lust,
  JHEP {\bf 0102}, 030 (2001)
  [hep-th/0012156];
%
  G.~Aldazabal, S.~Franco, L.~E.~Ibanez, R.~Rabadan and A.~M.~Uranga,
  JHEP {\bf 0102}, 047 (2001)
  [hep-ph/0011132].


\bibitem{Blumenhagen:2005mu}
  {\it For a reivew,}
  R.~Blumenhagen, M.~Cvetic, P.~Langacker and G.~Shiu,
  Ann.\ Rev.\ Nucl.\ Part.\ Sci.\  {\bf 55}, 71 (2005)
  [hep-th/0502005].





\bibitem{Camara:2003ku}
  P.~G.~Camara, L.~E.~Ibanez and A.~M.~Uranga,
  Nucl.\ Phys.\ B {\bf 689}, 195 (2004)
  [hep-th/0311241];
%
  D.~Lust, S.~Reffert and S.~Stieberger,
  Nucl.\ Phys.\ B {\bf 706}, 3 (2005)
  [hep-th/0406092];
%
  Nucl.\ Phys.\ B {\bf 727}, 264 (2005)
  [hep-th/0410074];
%
  L.~E.~Ibanez,
  Phys.\ Rev.\ D {\bf 71}, 055005 (2005)
  [hep-ph/0408064];
%
  A.~Font and L.~E.~Ibanez,
  JHEP {\bf 0503}, 040 (2005)
  [hep-th/0412150];
%
  E.~Floratos and C.~Kokorelis,
  hep-th/0607217.




\bibitem{Cremades:2003qj}
  D.~Cremades, L.~E.~Ibanez and F.~Marchesano,
  JHEP {\bf 0307}, 038 (2003)
  [hep-th/0302105];
%
  JHEP {\bf 0405}, 079 (2004)
  [hep-th/0404229].


\bibitem{Dutta:2005bb}
  B.~Dutta and Y.~Mimura,
  Phys.\ Lett.\ B {\bf 633}, 761 (2006)
  [hep-ph/0512171].


\bibitem{Dutta:2006bp}
  B.~Dutta and Y.~Mimura,
  Phys.\ Lett.\ B {\bf 638}, 239 (2006)
  [hep-ph/0604126].







\bibitem{Blanke:2006ig}
  M.~Blanke, A.~J.~Buras, D.~Guadagnoli and C.~Tarantino,
  JHEP {\bf 0610}, 003 (2006)
  [hep-ph/0604057];
%
  Z.~Ligeti, M.~Papucci and G.~Perez,
  Phys.\ Rev.\ Lett.\  {\bf 97}, 101801 (2006)
  [hep-ph/0604112];
%
  P.~Ball and R.~Fleischer,
  hep-ph/0604249;
%
  S.~Khalil,
  Phys.\ Rev.\ D {\bf 74}, 035005 (2006)
  [hep-ph/0605021];
%
  Y.~Grossman, Y.~Nir and G.~Raz,
  Phys.\ Rev.\ Lett.\  {\bf 97}, 151801 (2006)
  [hep-ph/0605028];
%
  F.~J.~Botella, G.~C.~Branco and M.~Nebot,
  hep-ph/0608100;
%
  G.~W.~S.~Hou,
  hep-ph/0611154.




\bibitem{Dutta:2006gq}
  B.~Dutta and Y.~Mimura,
  hep-ph/0607147.
  (To be published in Phys.\ Rev.\ Lett.)





\bibitem{D0}
D0 Collaboration, D0note 5144-Conf; D0note 5189-Conf;
\url{http://www-d0.fnal.gov/Run2Physics/WWW/results/b.htm}.



\bibitem{unknown:2006aq}
  B.~Aubert, {\it et al.}
  [BABAR Collaboration],
  hep-ex/0607107;
%
  K.~F.~Chen, {\it et al.}
  [Belle Collaboration],
  hep-ex/0608039.


\bibitem{Barberio:2006bi}
  E.~Barberio {\it et al.}  [Heavy Flavor Averaging Group (HFAG)],
  hep-ex/0603003;
  \url{http://www.slac.stanford.edu/xorg/hfag/.}





\bibitem{Ahuja:2006fv}
  G.~Ahuja, M.~Gupta, S.~Verma and M.~Randhawa,
  hep-ph/0608074.



\bibitem{ball}
  P.~Ball,
  hep-ph/0611108.



\bibitem{Kowalewski}
R.~Kowalewski, talk at ICHEP 06, Moscow, Russia;
  hep-ex/0610059.








\bibitem{Minkowski:1977sc}
  P.~Minkowski,
  Phys.\ Lett.\ B {\bf 67}, 421 (1977);
%
T. Yanagida, in proc. of KEK workshop, eds. O. Sawada and S. Sugamoto
(Tsukuba, 1979);
%
M. Gell-Mann, P. Ramond and R. Slansky,
in {\it Supergravity}, eds. P. van Nieuwenhuizen and D.~Z. Freedman
(North-Holland, Amsterdam, 1979);
%
  R.~N.~Mohapatra and G.~Senjanovi\'c,
  Phys.\ Rev.\ Lett.\  {\bf 44}, 912 (1980).



\bibitem{Apollonio:1999ae}
  M.~Apollonio {\it et al.}  [CHOOZ Collaboration],
  Phys.\ Lett.\ B {\bf 466}, 415 (1999)
  [hep-ex/9907037].



\bibitem{Schechter:1980gr}
  J.~Schechter and J.~W.~F.~Valle,
  Phys.\ Rev.\ D {\bf 22}, 2227 (1980);
%
  R.~N.~Mohapatra and G.~Senjanovic,
  Phys.\ Rev.\ D {\bf 23}, 165 (1981);
%
  G.~Lazarides, Q.~Shafi and C.~Wetterich,
  Nucl.\ Phys.\ B {\bf 181}, 287 (1981).







\bibitem{Brooks:1999pu}
  M.~L.~Brooks {\it et al.}  [MEGA Collaboration],
  Phys.\ Rev.\ Lett.\  {\bf 83}, 1521 (1999)
  [hep-ex/9905013].



\bibitem{Aubert:2005ye}
  B.~Aubert {\it et al.}  [BABAR Collaboration],
  %
  Phys.\ Rev.\ Lett.\  {\bf 95}, 041802 (2005)
  [hep-ex/0502032];
%
  Phys.\ Rev.\ Lett.\  {\bf 96}, 041801 (2006)
  [hep-ex/0508012];
%
  K. Abe {\it et al.}  [Belle Collaboration],
  hep-ex/0609049.




\bibitem{Bennett:2002jb}
  G.~W.~Bennett {\it et al.}, 
  Phys.\ Rev.\ Lett.\  {\bf 89}, 101804 (2002)
  [hep-ex/0208001];
%
  K.~Hagiwara, A.~D.~Martin, D.~Nomura and T.~Teubner,
  hep-ph/0611102.



\end{thebibliography}
\end{document}